\documentclass[aps,prb,twocolumn,showpacs,superscriptaddress,groupedaddress]{revtex4}
\usepackage{graphicx}  % needed for figures
\usepackage{dcolumn}   % needed for some tables
\usepackage{bm}        % for math
\usepackage{amssymb}   % for math

\usepackage{amsmath}   % math
\usepackage[colorlinks=true]{hyperref} % hyperlinks
\usepackage{upgreek}   % non-italic greek symbols

\begin{document}

\title{Spin noise of a polariton laser}

\author{I.~I.~Ryzhov}
\affiliation{Spin Optics Laboratory, St.~Petersburg State University, 1 Ul'anovskaya, Peterhof, St.~Petersburg 198504, Russia}

\author{M.~M.~Glazov}
\affiliation{Ioffe Institute, 26 Polytechnicheskaya, St.~Petersburg 194021, Russia}
\affiliation{Spin Optics Laboratory, St.~Petersburg State University, 1 Ul'anovskaya, Peterhof, St.~Petersburg 198504, Russia}

\author{A.~V.~Kavokin}
\affiliation{Spin Optics Laboratory, St.~Petersburg State University, 1 Ul'anovskaya, Peterhof, St.~Petersburg 198504, Russia}
\affiliation{Department of Physics \& Astronomy, University of Southampton, Southampton SO17 1BJ, United Kingdom}
\affiliation{CNR-SPIN, Viale del Politecnico 1, I-00133 Rome, Italy}

\author{G.~G.~Kozlov}
\affiliation{Spin Optics Laboratory, St.~Petersburg State University, 1 Ul'anovskaya, Peterhof, St.~Petersburg 198504, Russia}

\author{M.~A\ss{}mann}
\affiliation{Experimentelle Physik 2, Technische Universit\"{a}t Dortmund, D-44221 Dortmund, Germany}

\author{P.~Tsotsis}
\affiliation{Microelectronics Research Group, IESL-FORTH, P.O. Box 1385, 71110 Heraklion, Greece}

\author{Z.~Hatzopoulos}
\affiliation{Microelectronics Research Group, IESL-FORTH, P.O. Box 1385, 71110 Heraklion, Greece}

\author{P.~Savvidis}
\affiliation{Department of Materials Science and Technology, University of Crete, P.O. Box 2208, 71003 Heraklion, Greece}
\affiliation{Microelectronics Research Group, IESL-FORTH, P.O. Box 1385, 71110 Heraklion, Greece}

\author{M.~Bayer}
\affiliation{Experimentelle Physik 2, Technische Universit\"{a}t Dortmund, D-44221 Dortmund, Germany}
\affiliation{Ioffe Institute, 26 Polytechnicheskaya, St.~Petersburg 194021, Russia}

\author{V.~S.~Zapasskii}
\affiliation{Spin Optics Laboratory, St.~Petersburg State University, 1 Ul'anovskaya, Peterhof, St.~Petersburg 198504, Russia}

\begin{abstract}
	We report on experimental study of the exciton-polariton emission (PE) polarization noise below and above the polariton lasing threshold under continuous wave non-resonant excitation. The experiments were performed with a high-Q graded $5\lambda/2$ GaAs/AlGaAs microcavity with four sets of three quantum wells  in the strong coupling regime. The PE polarization noise substantially exceeded in magnitude the shot noise level and, in the studied frequency range (up to 650 MHz), had a flat spectrum. We have found that the polarization and intensity noise dependences on the pump power are strongly different. This difference is ascribed to the bosonic stimulation effect in spin-dependent scattering of the polaritons to the condensate. A theoretical model describing the observed peculiarity of the PE polarization noise is proposed.
\end{abstract}

\maketitle

\emph{Introduction.} Nowadays, the term ``laser'' is applied to any device producing coherent, monochromatic and unidirectional light~\cite{1}. It turns out that stimulated emission of radiation is not the only way to generate laser light. In \textit{polariton lasers}, light is emitted spontaneously by a condensate of bosonic quasiparticles, exciton-polaritons, accumulated in a single quantum state~\cite{2,TS,Chrs,3}. Polariton lasers do not require an exciton-polariton population inversion and the emission may result from the quasi-equilibrium ensemble of exciton-polaritons.
In a polariton laser, a semiconductor microcavity is excited non-resonantly, either optically or electrically. A gas of electrons and holes created in the cavity forms excitons, which subsequently thermalize, mainly through exciton-exciton interactions. Their kinetic energy is lowered by interactions with phonons and they relax along the lower polariton dispersion branch. The polaritons finally scatter to their lowest energy state, where they accumulate because of the Bose-stimulation. The coherence of the particles therefore builds up from an incoherent reservoir~\cite{TS,3,belykh}. The polariton lasers emit light due to photon tunneling through the Bragg mirrors. This emission is spontaneous; however, the light going out has all properties of a laser light: It is coherent, monochromatic, polarized and unidirectional. 
The stimulated scattering of polaritons and polariton lasing have been realized in planar and micropillar GaAs, CdTe, GaN microcavities~\cite{S5,S5-1,Chrs,7}. Room temperature operation has been demonstrated in GaN~\cite{Chrs1} and ZnO-based~\cite{ZnO,ZnO1} polariton lasers. Recently, polariton lasers with electrical injection of carriers have been realized~\cite{4,4-1}. This technological breakthrough opens way to a new generation of optoelectronic devices based on the Bose-Einstein condensates of mixed light-matter quasiparticles. This fact attracts great interest to further research of fundamental properties of polariton emitters. Along with investigations of regular properties of such systems, like small-signal modulation characteristics of an electrically pumped polariton laser~\cite{9}, a great amount of important physical information is provided by studying  their stochastic properties. It should be noted that while the statistics of exciton-polaritons in polariton condensates has been widely studied through the second order coherence measurements~(see e.g., Refs.~\onlinecite{Jean, Kim}), the spin noise in polariton condensates has not been studied experimentally to the best of our knowledge. At the same time, specific parameters of spin noise are crucial for applications, on one hand, and provide fundamental understanding of the emitting state nature, on the other. Theoretically, a giant polarization noise is expected in the polariton lasing regime~\cite{10}. The origin of the giant noise is in the stochastic formation and bosonic amplification of polarization of exciton-polariton condensates~\cite{11,12}.
Here we study  the polariton emission (PE) noise in a quantum-well microcavity above the polariton lasing threshold under continuous wave (\emph{cw}) excitation. Main attention is paid to the polarization noise of the emission, carrying most specific information about dynamics of the polariton formation, relaxation, as well as of their interactions. Polarization noise is characterized by strong magnitude and broad bandwidth. A strongly non-monotonic dependence of the PE polarization noise power on the pump intensity are found and interpreted.

\emph{Experimental.} In contrast to standard measurements in spin noise spectroscopy (see, e.g., Refs.~\onlinecite{13,13-1,13-2,14}), in this work we analyzed polarization fluctuations of secondary emission, rather than of a probe beam. 
\begin{figure}
 \includegraphics[width=0.95\columnwidth,clip]{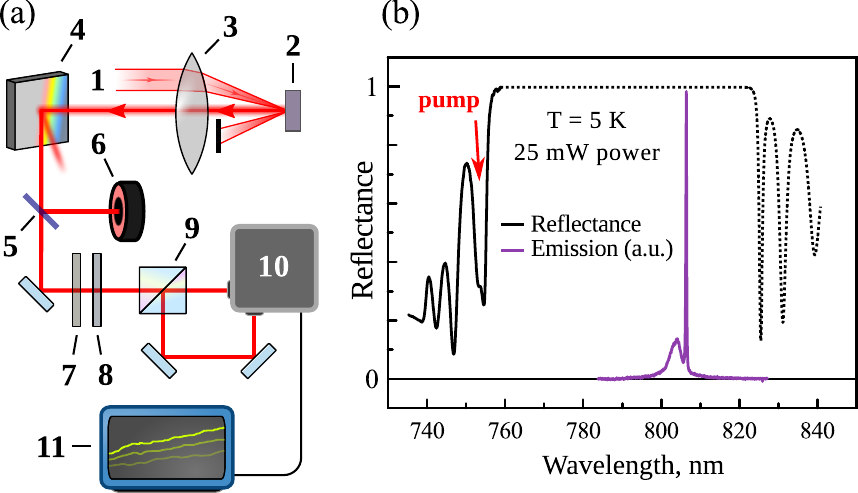}
 \caption{(a) Schematics of the experimental setup (see description in the text). (b) Reflection (black) and photoluminescence (purple) spectra of the microcavity structure. Solid curves represent the measured spectra and the dashed line is a schematic representation of sample's outer DBR reflectivity. }
 \label{fig:setup}
\end{figure}
The experiments were performed with a high-Q graded  $5\lambda/2$ GaAs/Al$_{0.3}$Ga$_{0.7}$As microcavity containing four sets of three 10-nm GaAs quantum wells, located at the antinodes of the cavity, and Bragg mirrors formed of Al$_{0.15}$Ga$_{0.85}$As/AlAs layers (see Ref.~\onlinecite{15} for details). Strong exciton-photon coupling provided opportunity to selectively detect emission from the lower polariton branch. Figure~\ref{fig:setup} shows schematics of  the experimental setup (a) and the reflectivity and emission spectra of the sample (b). Output emission of a \emph{cw} Ti:sapphire laser (\textbf{1}) was focused on the sample (\textbf{2}) by a Helios-44M objective (\textbf{3}). The beam was incident on the sample at a small angle to remove reflected light from the detector. The wavelength of the pump light was chosen to match the nearest to the stop-band reflectivity dip at about 755 nm. The diameter of the light spot on the sample was about 15~$\mu$m. Position of the spot on the sample was chosen to provide the most efficient polariton emission, which corresponded to small negative detunings.

The light emitted from the sample along its growth direction was sent to diffraction grating (\textbf{4}) to filter out the scattered pump light. A non-polarizing beam splitter (\textbf{5}) diverts a fraction of the light beam to a powermeter (\textbf{6}). The rest of the light beam is directed to the conventional spin noise detection system (see, e.g., Ref.~\onlinecite{14}) comprised of the polarization beam splitter (\textbf{9}) and the balanced detector (\textbf{10}). The output signal of the latter is fed to the FFT spectrum analyzer (\textbf{11}). We also used additional (half-wave and quarter-wave) phase plates (\textbf{7}) and (\textbf{8}) mounted in front of the beam splitter. The PE {\it intensity} noise was measured using a single detector of the balanced scheme without any preliminary polarization analysis. The bandwidth of the detectors used in this setup was either 100 or 650 MHz. The noise power spectrum was essentially flat up to $\gtrsim 650$~MHz, hence, in our measurements we integrated the signal of spectrum analyzer over a frequency range of 100 MHz to improve the measurements accuracy.

\begin{figure}
 \includegraphics[width=.95\columnwidth,clip]{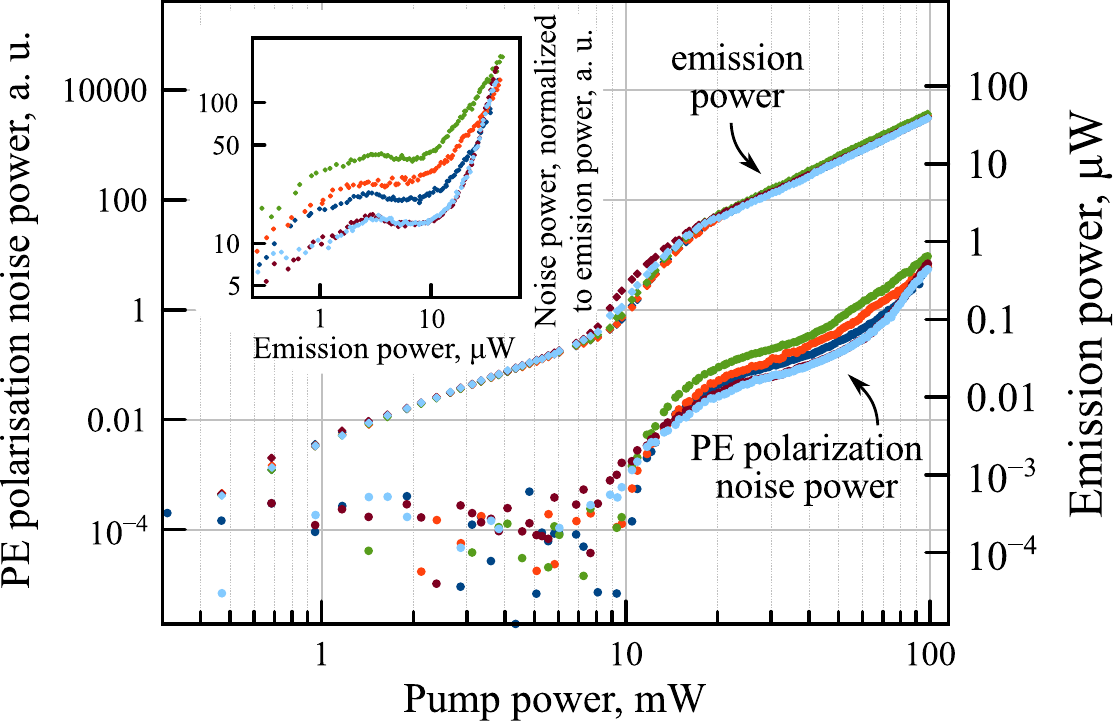}
 \caption{Typical dependences of the PE power and polarization noise power on the pump intensity obtained for several randomly chosen points on the sample. Inset represents the dependence of polarization noise power normalized to the PE intensity on the PE intensity itself.}
 \label{fig:emission-power}
\end{figure}

\emph{Results and discussion}.
The sample was found to be rather inhomogeneous: changing position of the objective by a few microns could substantially change characteristics of the PE. The emission spectrum typically contained several narrow peaks in the region of the lower polaritonic branch evolving independently with the pump power. Hence, in most cases, we detected the light from several emitters. This was implicitly revealed in power dependence of the noise magnitude (see below). 
In spite of this inhomogeneity, the dependence of the PE intensity on the pump power was qualitatively the same for all spots on the sample, Fig.~\ref{fig:emission-power}: In the region of low intensities, the PE power increased linearly, then, in the vicinity of 10 mW, it exhibited a sharp bend indicating threshold of polariton lasing, after which it showed a monotonic superlinear growth with practically no essential features. 

Polarization of the PE, in our experimental setup, was distorted by the diffraction grating, but since it did not show any noticeable birefringence, it could only slightly tilt the polarization plane or polarization ellipse. On the basis of polarization measurements with a quarter-wave plate installed in the pump beam, we have found that under linearly (or circularly) polarized pump, the PE was partially linearly (or, respectively, circularly) polarized.

The PE polarization noise proved to be fairly strong and covering a wide frequency range extending far beyond the bounds of bandwidth of our detection system. At the same time it can be shown that, under our experimental conditions, for the width of the polarization noise spectrum 10\ldots100 GHz, the ratio of power density of the polarization noise to that of the shot noise should be 1--2 orders of magnitude higher than what is observed experimentally. It agrees with the assumption that, generally, the detected light is contributed by many emitters, thus reducing instantaneous degree of polarization.

Bearing in mind the description of polarized light using the Stokes vector $\bm S = (S_x, S_y, S_z)$, one can see that the light polarization fluctuations can be described just like the electron spin noise by the set of correlation functions of the Stokes vector components~\cite{10,glazov:noneq,note:class}. The light unpolarized on average  may result from fluctuations of the polarization plane azimuth or from fluctuations of the light ellipticity. We performed additional measurements of ellipticity noise amplitude using quarter-wave plate in the detection channel and found that the predominantly linearly or circularly polarized light (controlled by the pump polarization) mainly shows, respectively, fluctuations of the polarization plane azimuth or ellipticity.  

Typical dependences of the PE polarization noise power on the pump power are shown in Fig.~\ref{fig:emission-power}. This plot reveals the threshold build-up in a more pronounced way than the PE power itself. After the threshold jump, all the curves show the same behavior with vague traces of peculiarities, which are revealed better once the polarization noise power is normalized to the PE intensity (see inset in Fig.~\ref{fig:emission-power}).

\begin{figure}
 \includegraphics[width=0.95\columnwidth,clip]{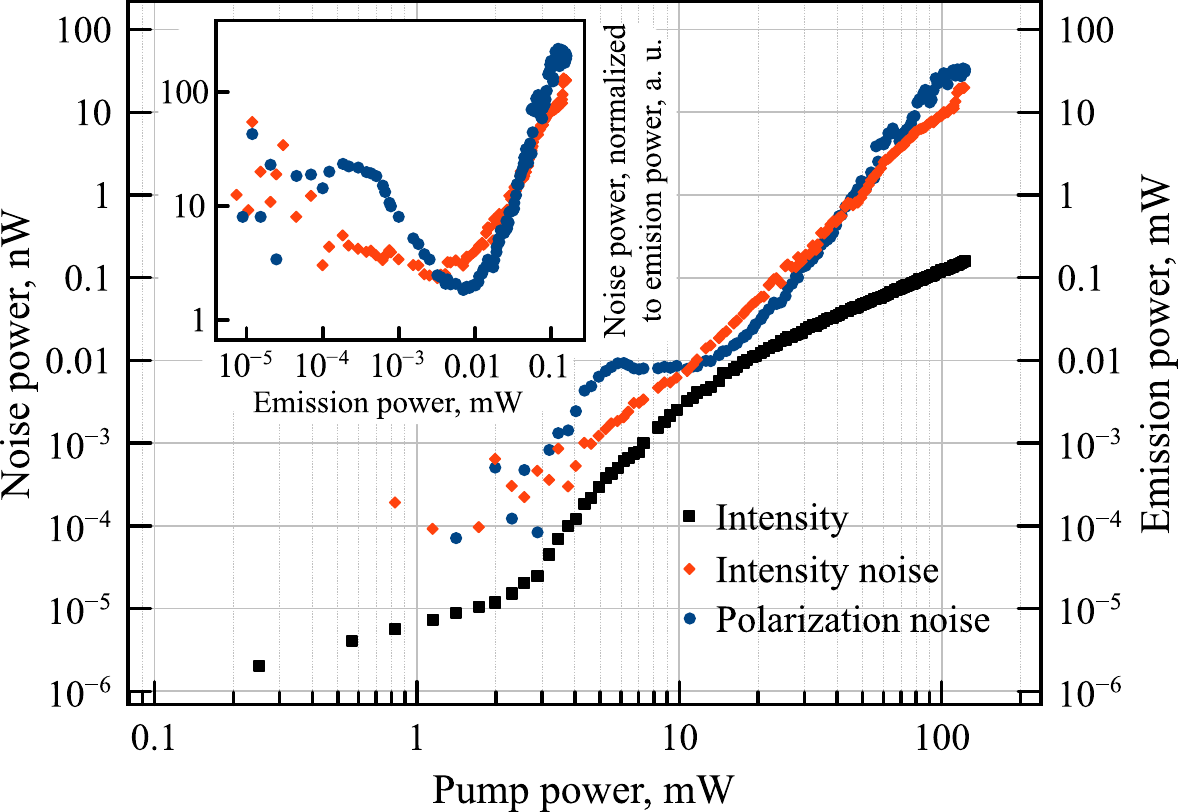}
 \caption{An example of dependences of PE intensity (black squares), PE intensity noise (red rhombuses), and PE polarization noise (blue circles) on the pump power obtained at the same point on the sample. Inset shows intensity (red rhombuses) and polarization (blue circles) noise normalized to the emission intensity.  }
 \label{fig:int-pol}
\end{figure}

\begin{figure}
 \includegraphics[width=0.85\columnwidth,clip]{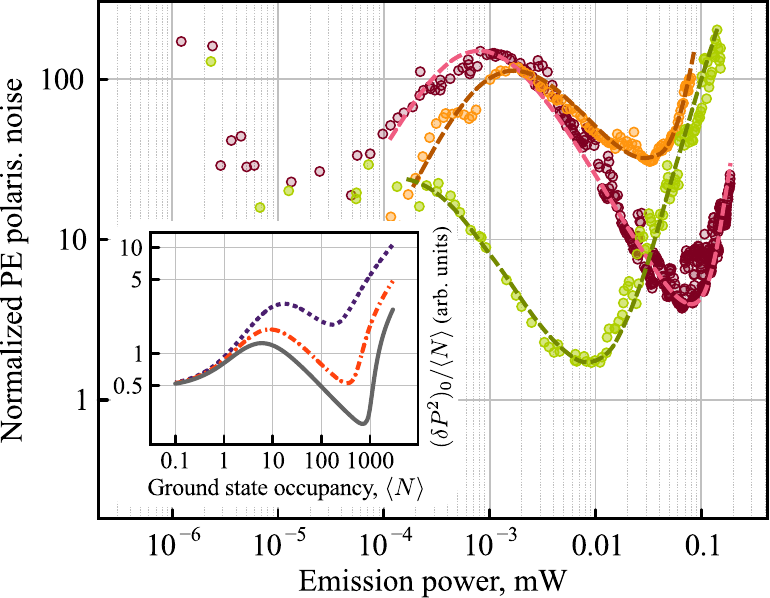}
 \caption{Several (specially chosen) dependence of the normalized PE polarization noise power on the emission intensity. Dashed lines are guides to the eye. The inset represents normalized polarization noise as a function of the ground state occupancy calculated for $g^{(2)}=1$ and $\alpha/\Upgamma_0=0.025$ (violet/dotted), $0.05$ (red/dash-dotted), $0.075$ (gray/solid). To model the transition to the photon lasing, we assumed that the interaction constant smoothly vanishes in the range of $\langle N \rangle$ from 100 to 200 (violet/dotted), 400 to 600 (red/dash-dotted) and 900 to 1100 (gray/solid). }
 \label{fig:normalized-pn}
\end{figure}

At some points on the sample, however, such peculiarities could be observed in a much more spectacular form, as in Fig.~\ref{fig:int-pol}, where, the PE intensity (black squares), intensity noise (red rhombuses), and linear polarization noise (blue circles) are shown for the same point on the sample. We see that a strongly pronounced feature on the curve of the polarization noise is not revealed in two other dependences. This fact is additionally illustrated by the power dependences of the PE polarization and intensity noise (both normalized to the PE power) shown in the inset of Fig.~\ref{fig:int-pol}. These normalized dependences allow one to reveal clearly these universal peculiarities of the PE polarization noise. Figure~\ref{fig:normalized-pn} shows several dependences of the intensity-normalized polarization noise power on the PE power for a few specially chosen points on the sample. We see that these dependences are systematically non-monotonic, showing peaks at some level of excitation and subsequent decrease of the polarization noise with the PE power, followed by the growth at high emission/excitation powers. An interplay of individual emitters with different non-monotonic contributions to the detected PE may smoothen the dependences.
%structure of the curves in Fig.~\ref{fig:normalized-pn}.

The main features of the PE polarization noise and, especially, peculiarities of its power dependence can be qualitatively understood following the general model of polariton spin noise developed in Ref.~\onlinecite{10}. Firstly, due to the short polariton lifetime ($\sim 1\ldots 10$~ps), which was evaluated in Ref.~\onlinecite{15}, the intensity and polarization noise spectra are flat in the addressed frequency range, and it is sufficient to consider the noise contribution at zero frequency. For the polariton ensemble with $\langle N \rangle$ particles in the ground state (emission intensity $I \propto \langle N \rangle$) the intensity- and spin-noise power can be presented, respectively, as~\cite{10}
\begin{equation}
\label{noise}
(\delta I^2)_0 \propto \frac{\langle \delta N^2 \rangle}{\Upgamma_N}, \quad (\delta P^2)_0 \propto \frac{\langle \delta N^2 \rangle}{\Upgamma_S}.
\end{equation}
Here $\langle \delta N^2 \rangle$ is the mean square fluctuation of the particle number in the ground state and $\Upgamma_N$ and $\Upgamma_S$ are the decay rates for fluctuations of the particle number and the Stokes vector components, respectively. Hereafter, we consider the fluctuations of the linear polarization which directly correspond to the fluctuations of the in-plane Stokes vector components, $S_x$, $S_y$~\cite{10}. An increase of the pumping rate gives rise to increase of the ground state occupancy, $\langle N \rangle$, and of the mean square fluctuations. The latter is determined by the ground states statistics, $g^{(2)}$:
\begin{equation}\label{eq:ms-fluct}
\langle \delta N^2 \rangle = \frac{1}{2} \langle N\rangle [1 + (g^{(2)} - 1) \langle N \rangle].
\end{equation}
For the coherent statistics relevant for polariton lasers operating above the threshold~\cite{19} $g^{(2)} = 1$ and $\langle \delta N^2 \rangle \sim \langle  N \rangle$. The decay of the particle number is governed by an interplay of the photon decay through the mirrors and the stimulated scattering towards the ground state. As a result, the fluctuations are supported by the stimulated scattering and $\Upgamma_N = \Upgamma_0/(1 + \langle N \rangle)$, where $\Upgamma_0$ is the polariton decay rate in the linear regime~\cite{10}. As a result, for the intensity noise normalized to the ground state occupancy (i.e. to the emission intensity) well above the threshold, $\langle N \rangle \gg 1$, and $g^{(2)}=1$ one has
\begin{equation}
\label{i:noise}
(\delta I^2)_0/\langle N\rangle \propto \langle N\rangle/{\Upgamma_0}
\end{equation}
For the Stokes vector fluctuations decay rate one has~\cite{10}
%
%\begin{equation}\label{eq:Gamma-s}
$\Upgamma_S = \Upgamma_N + \upgamma_S,$
%\end{equation}
%
where $\upgamma_S$ is the spin decoherence rate. For the in-plane Stokes vector components responsible for the linear polarization of emission an efficient channel of the decoherence is the self-induced Larmor precession~\cite{17,18} in the effective field caused by fluctuating $S_z$ component. For $\langle N \rangle \gg 1$ we obtain~\cite{10}
\begin{equation}\label{eq:gamma-s}
\upgamma_S = |\alpha| \sqrt{\langle \delta N^2 \rangle},
\end{equation}
where $\alpha$ is the effective constant of the polariton-polariton interaction  for the parallel spin configuration; the interactions of polaritons with opposite spins are neglected and the numerical coefficient is included in $\alpha$. As a result, for the normalized polarization noise we get
\begin{equation}
\label{p:noise}
(\delta P^2)_0/\langle N\rangle \propto \frac{\langle N\rangle}{{\Upgamma_0}+|\alpha|\langle N\rangle^{3/2}}.
\end{equation}

The spin decoherence rate increases with increasing mean occupancy of the ground state, Eq.~\eqref{eq:gamma-s}. It results in drastically different behavior of the intensity and linear polarization noise: According to Eq.~\eqref{i:noise} the normalized intensity noise grows monotonously with the $\langle N \rangle$, while the normalized noise of polarization $(\delta P^2)_0/\langle N \rangle$ behaves non-monotonously with $\langle N \rangle$. First, it increases due to the decrease of $\Upgamma_S$ as a result of Bose-stimulation and corresponding decrease of the first term  in the denominator of Eq.~\eqref{p:noise}. Then, the interactions become sufficiently strong, and the second term in the denominator of Eq.~\eqref{p:noise} starts to dominate, and the normalized polarization noise, $(\delta P^2)_0/\langle N \rangle$, decreases. The further increase in the pump intensity may lead to the photon lasing, when the strong coupling is lost. In this case, the interactions play a minor role and, consequently, the polarization and intensity fluctuations behave similarly. That is why the normalized polarization noise grows with the further increase of $\langle N \rangle$. It is illustrated in the inset of~Fig.~\ref{fig:normalized-pn} where the results of the calculations are shown. 

We stress that the presented model provides just a possible scenario, which qualitatively explains the experimental data. Due to the substantial inhomogeneity of the sample, the quantitative agreement is not possible in this simple model. Moreover, other factors such as the dependence of the polariton second-order coherence, $g^{(2)}$, on the pump power~\cite{19}, anisotropic splitting of the polariton states, and the effect of interactions on the magnitude of pseudospin-$z$ component and, correspondingly, on the ellipticity fluctuations~\cite{10} should be taken into account in  realistic modeling. Additionally, repulsive interparticle interactions may also suppress particle density fluctuations  because such fluctuations are energetically unfavorable~\cite{kagan}. This might explain a slight reduction of the intensity noise (see inset in Fig.~\ref{fig:int-pol}). We do not address here the fluctuations below the threshold since the emission signals are very weak in this region.

The threshold-like increase in spin fluctuations, well correlating with our results, has been reported in Refs.~\onlinecite{16,sala}. Unlike our work, Refs.~\onlinecite{16,sala} used \emph{pulsed} excitation and detected momentary polarizations corresponding to the peak intensities. Hence, the results of Ref.~\onlinecite{16} correspond to the momentary fluctuations $\langle \delta S_\alpha^2\rangle$ rather than to the zero-frequency spin noise power, Eq.~\eqref{noise}. Moreover, the non-equilibrium fluctuations are not related to any susceptibility of the system~\cite{gants,glazov:noneq}, and there is no direct correlation between the \emph{cw} values and the pulsed response. A further insight into the  spin noise of polaritons requires measurements of spin noise spectra in the whole range up to $10^2\ldots 10^3$~GHz, e.g., by the ultrafast spin noise spectroscopy~\cite{oest:uf}.

\emph{Conclusion.} In this work, we studied polarization noise of polariton emission of a quantum-well microcavity under \emph{cw} excitation in the vicinity of the lasing threshold. In contrast to standard conditions of the spin noise spectroscopy, we had to detect, in this case, strong polarization noise of a weakly polarized light, rather than small polarization fluctuations of a perfectly polarized beam. We have discovered that variations of the polarization noise with power and polarization of the pump reveal specific features hidden in the conventional intensity-related properties of the polariton emission. For instance, the polarization noise is apparently a much more sensitive quantity for mapping thresholds than the intensity or the intensity noise. We believe that these new opportunities of research provided by the polarization-noise technique  will allow one in future to obtain important information, inaccessible for other methods, about dynamics of the polariton Bose-condensate under \emph{cw} excitation. 

\emph{Acknowledgements.} Financial support from the SPbGU (Grant No.~11.38.277.2014), the RFBR and DFG in the frame of International Collaborative Research Center TRR 160 (project No.~15-52-12013) is acknowledged. The work is supported by RFBR grant No.~16-52-150008. IR acknowledges RFBR project No.~16-32-00593 and by EU FP7-REGPOT-2012-2013- 1 Grant No. 316165. MG also acknowledges partial support by the RFBR project No.~14-02-00123, the Russian Federation President Grant MD-5726.2015.2 and the Dynasty foundation. AK acknowledges the fellowship from EPSRC Established Career (Grant No. RP008833). The work was carried out using the equipment of SPbU Resource Center `Nano\-pho\-tonics' (photon.spbu.ru).


\begin{thebibliography}{100}
\bibitem{note:class} Here the light field can be considered as a classical wave, otherwise field operators should be properly ordered~\cite{carm}.
\end{thebibliography}
\end{document}